\pdfoutput=1

\documentclass{acm_proc_article-sp}

\usepackage[utf8]{inputenc}
\usepackage{subcaption}
\usepackage{booktabs}
\usepackage[T1]{fontenc}
\usepackage{times}
\usepackage{helvet}
\usepackage{courier}
\usepackage[russian,turkish]{babel}
\usepackage{dcolumn}
\usepackage{balance}

\makeatletter
\def\@copyrightspace{\relax}
\makeatother

\begin{document}

\shorthandoff{=}

\title{Dawn of the Selfie Era: The Whos, Wheres, and Hows of Selfies on Instagram}

\numberofauthors{1}
\author{
  \alignauthor Flávio Souza$^\dag$~~~~~~~Diego de Las Casas$^\dag$~~~~~~~Vinícius Flores$^\dag$~~~~~~~SunBum Youn$^*$\\Meeyoung Cha$^*$~~~~~~~~Daniele Quercia$^*$~~~~~~~~~Virgílio Almeida$^\dag$\\
    \affaddr{$^\dag$Computer Science Department, UFMG, Belo Horizonte, Brazil}\\
	\affaddr{$^*$Graduate School of Culture Technology, KAIST, South Korea}
}

\maketitle

\begin{abstract}
Online interactions are increasingly involving images, especially those containing human faces, which are naturally attention grabbing and more effective at conveying feelings than text. To understand this new convention of digital culture, we study the collective behavior of sharing \textit{selfies} on Instagram and present how people appear in selfies and which patterns emerge from such interactions. Analysis of millions of photos shows that the amount of selfies has increased by 900 times from 2012 to 2014. Selfies are an effective medium to grab attention; they generate on average 1.1--3.2 times more likes and comments than other types of content on Instagram. Compared to other content, interactions involving selfies exhibit variations in homophily scores (in terms of age and gender) that suggest they are becoming more widespread. Their style also varies by cultural boundaries in that the average age and majority gender seen in selfies differ from one country to another. We provide explanations of such country-wise variations based on cultural and socioeconomic contexts. 
\end{abstract}

\category{J.4}{Computer Applications}{Social and Behavioral Sciences}
\category{H.3.5}{Information Storage and Retrieval}{Online Information Services}[Web-based services]
\terms{Human Factors; Measurement}
\keywords{Instagram; Selfies; Cultural Boundaries}

\section{Introduction}

The amount of rich media is increasing exponentially on the Internet. Online conversations and interactions now involve more images, which are naturally attention grabbing and effective at conveying feelings~\cite{joo2014visual, winston2013photography}. Social media in particular has seen a rapid uptake of pictures containing human faces. One notable example is the \textit{selfie} or digital self-portrait, which have become a phenomenal ubiquitous convention of online culture. 

Numerous research studies proposed the psychological and sociological framing behind posting selfies, broadly based on narcissism~\cite{twenge2009narcissism}, self-exploration~\cite{schwarz2010friendship}, self-embellishment~\cite{marwick2015instafame}, and a new genre of art~\cite{winston2013photography}. Other studies approached with the Human Computer Interaction framing to understand pictures with faces and demonstrated their engaging effects~\cite{facesengageus}. In addition, a project called Selfiecity examined the image traits of single-person self-portraits in five cities across the world~\cite{tifentaleselfiecity}. Until now, little effort has been made to quantitatively defining and examining selfies based on a large amount of data.

This paper presents a measurement study of a popular media sharing website, Instagram (www.instagram.com), and characterizes {how people appear on Instagram selfies and which patterns emerge from their attention grabbing behaviors}. Since selfies are pictures of people, they represent a structured (i.e., social-by-design) form of interaction in social networks. We hence seek to understand whether this new content type can uncover  patterns of social interactions. We ask the following two specific questions. 

\begin{enumerate}
    \item \textbf{The whos and wheres of selfies:} Can we characterize selfies in terms of age, gender, geography, country, and other cultural variables?
    \item \textbf{The hows of selfies:} How much attention do selfies receive in terms of likes and comments and to what extent their interactions depend on cultural boundaries?
\end{enumerate}

The first question provides a holistic understanding of what selfies represent in social media. We utilize a subset of photos with hashtags containing the word `selfie' to determine what kinds of photos are explicitly called as selfies by Instagram users (e.g., how many persons appear in a photo and what kinds of moods these photos contain). Several critical hypotheses related to gender empowerment, group membership, and perceived privacy are tested to better understand the contexts through which users post selfies in a given culture.

Through the second research question, we try to understand how selfie users interact with their audience. Selfies and pictures with faces are more than mere self-expressions; they are phenomenal in grabbing attention  and have settled as a popular online practice. By studying the dyadic relationships between selfie users and their audience, we aim to understand what principles rule in pair-wise interactions that involve rich media content. This study tests whether conventional theories such as homophily become strengthened or weakened under the new form of interaction among users. 

This paper utilizes a large amount of data gathered from Instagram and carefully selected data about selfies\footnote{Data used in this study are available for research purposes at {http://instagram.camps.dcc.ufmg.br/selfies/}} based on three different approaches: (i) pictures containing the word `selfie' or its immediate variations in hashtags (e.g., \#selfie, \#myselfie), (ii)  pictures containing hashtags related to selfie, but composed only of indirect variations of the word (e.g., \#selfcamera, \#me), and (iii) pictures containing one or more faces, irrelevant to the choice of hashtags. In this manner, findings in this paper are not  dependent on a particular definition of selfies but can provide a holistic view of what people consider selfies. We make the following observations, which are explained throughout the paper.

\begin{enumerate}
    \item The amount of selfies increased by 900 times over 3 years from 2012 to 2014, which indicates the phenomenon has become a truly ubiquitous convention. 
    
	\item Selfies are effective in grabbing attention in social media; they receive 1.1--3.2 times more likes and comments from audience than general posts on Instagram.    

    \item  Young females are the most prominent group who appear in selfies around the world, except for certain countries such as Nigeria and Egypt that show male dominance. 
    
    \item There is a complex relationship between taking selfies and a country’s culture. The chance of using selfie-related hashtags was higher for cultures with stronger local community membership as well as weaker perception of privacy.

    \item Beyond cultural boundaries, selfie-based interactions present homophily variations over time in terms of both age and gender, suggesting that selfies are becoming more mundane.
\end{enumerate} 

This work contributes towards better understanding selfies as a popular online phenomenon that have evolved beyond fads, becoming an effective medium of interaction that is attention grabbing and increasing in demand. Our findings demonstrate that selfies are a new window to study collective user behaviors, providing important insights into  subjects like perception of privacy, digital cultural norms, and designs of social-networking platforms.

\section{Background}

The rise of selfies is a key trend in the visual Web, assisted by new technological tools and services like Flickr, Pinterest, and Instagram that allow people to better express themselves visually. This section describes several findings from research on self-portrait images, selfies, and Instagram.

\subsection{The Meaning of Selfies}
Selfies are a ubiquitous phenomenon of modern digital culture. The term was added to the Oxford Dictionaries\footnote{http://www.oxforddictionaries.com} in 2013, with description: a photograph that one has taken of oneself, typically one taken with a smartphone or webcam and shared via social media.  

Different theories emerged to explain why people take selfies. Some state selfies are a mean of self-exploration. As one takes multiple selfies and combine them with different filters, one can re-see herself~\cite{wp:knowthyselfie}. A slightly different view is self-embellishment from psychology that states when exposed to slightly modified pictures of themselves, people tend to identify a more attractive version as the original picture~\cite{wp:sciencebehind}. With the ability to control  aesthetics of a picture, selfies are a perfect tool for showing the world one's subjective self-image.

A sociological framing recognizes technological possibility to be a necessary condition and also highlights other behavioral factors to be important for selfies~\cite{wp:whyweselfie}. One is a culture of sharing and belonging fostered by the online environment and transmitted through memes. Another is the constant work of shaping and reaffirming self-identity through social actions. In this perspective, selfies could symbolize a convention that is governed by culture and society.

\subsection{Advocates and Opponents of Selfies}
Selfies are a prominent online culture that have been both criticized and advocated by different parties. Critics say selfies are vain, narcissistic, and attention-seeking; some argue a wide adoption of selfies by female users exacerbates sexual objectification and male gaze~\cite{wp:selfiedabates1}. One research demonstrated that adults with the Dark Triad personality trait (e.g., narcissism, psychopathy, and machiavellianism) have a higher chance of posting selfies and editing images on social media~\cite{fox2015}. Self-objectification is also known to correlate with increasing photo sharing activities on Facebook among young women~\cite{meier2014}.  This leads to a worry about the loss of control over one's self-image in an increasingly sharing and hackable culture, where the notion of privacy becomes dependent on the types of interactions that are allowed~\cite{privacycollective}. The mere presence of an individual's face in a public photo stream can reveal a great detail of information about that person~\cite{estimatingheights}.

Defenders of the selfie culture not only deny the above claims but argue selfies are the pinnacle of control and self-expression; selfies allow people to take control over how they and their peers are represented in public, which mobilizes the power dynamics of representations and promotes empowerment~\cite{wp:selfiedebates2}. One study interviewed 20 participants who had posted sexual self-portraits and showed how the exchange of such self-portraits can be a transformative experience, increasing their critical self-awareness in a positive manner~\cite{tiidenberg2014}. 

\subsection{Selfies by Numbers}
In contrast to the rich body of work on sociological interpretation of selfies, relatively little attention has been given to data-driven analysis of selfies. A report by eBay Deals states that selfie activity is platform dependent and is well distributed in particular media, for instance Instagram than Twitter~\cite{wp:selfierevolution}. A research conduced by TIME looked at how many ``selfies per capita'' each city produced by dividing the amount of users posting selfies by the population of each city. They noticed that it was difficult to find a proper local translation for the hashtag `selfie', as different variations were used everywhere~\cite{timeselfiecapital}. 

The largest scale analysis of selfies to date, however, probably was a data visualization project called Selfiecity that aimed at describing features of single selfies (i.e., photos containing a single person's face) in five cities across the world~\cite{tifentaleselfiecity}. They investigated demographics, poses, face features, and the moods of 3,200 selfies on Instagram using both automatic and manual methods. Nonetheless, many of the considerations and theories behind selfies (e.g., the contexts, interactions) have not yet been studied under the perspective of data analysis, which is the goal of this paper.

\subsection{Studies on Instagram}
When it comes to general user behaviors, a number of research utilized the logs gathered from  Instagram. For example, researchers examined how color patterns varies between photos posted in two cities~\cite{hochman2012visualizing}, how the behaviors of teens and adults on the network differ~\cite{generationlike}, how users can be grouped based on the types of content they share~\cite{whatweinstagram}, and even how Instagram photos shared on Twitter can be used as sensors to study user characteristics in different cultures~\cite{thousandwords}. Therefore, the present research can be seen yet as an additional contribution for Instagram characterization efforts, complementing previous works in this direction.

\section{Instagram Data}

We started data collection by inferring ranges of user IDs. This step involved forward sampling batches of 10,000 numeric IDs for every range of 10 million, starting from zero. None of the inspected IDs were valid after the count of 1.6 billion. Through this process, we could identify which specific ranges are valid ID space. Based on these ranges, we next randomly sampled 1\% or 16 million IDs to build an initial seed set and found 42\% of them to be in use; the remaining IDs were either deleted or not in use. Not all of these in-use accounts could be viewed publicly due to privacy settings;  78\% of them were public accounts and the remaining 22\% were private accounts, whose profile and feed information could be viewed only by  confirmed friends on Instagram.  

\begin{figure}[h]
    \centering
    \begin{subfigure}[t]{0.45\columnwidth}
        \centering
        \includegraphics[width=\columnwidth]{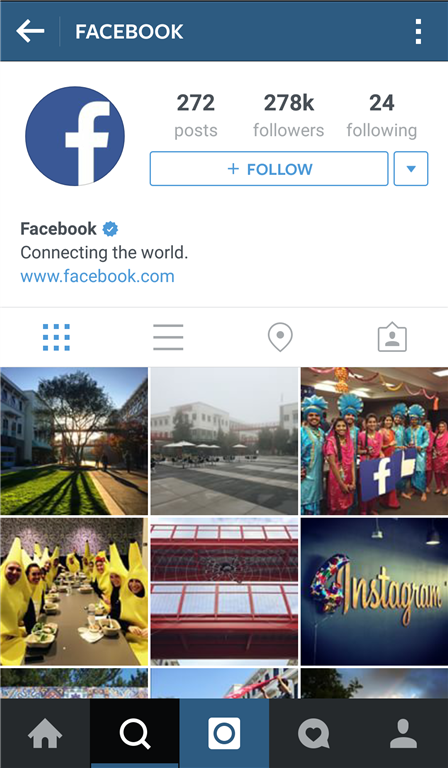}
        \caption{Profile}
        \label{fig:insta-user}		
    \end{subfigure}
    \qquad
    \begin{subfigure}[t]{0.45\columnwidth}
        \centering
        \includegraphics[width=\columnwidth]{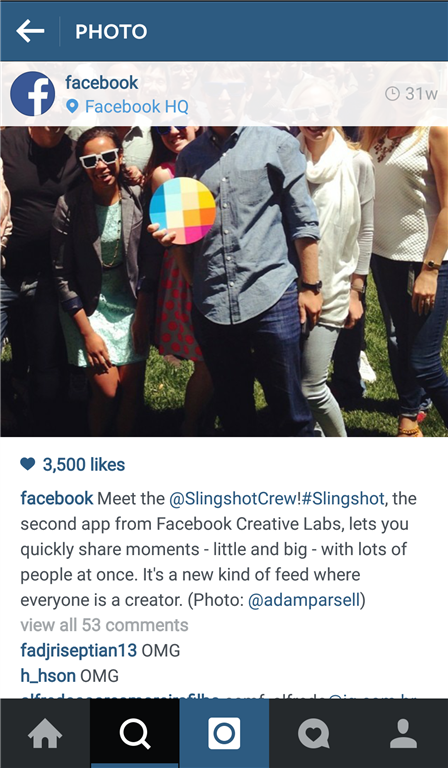}
        \caption{Feed}
        \label{fig:insta-media}
    \end{subfigure}
    \caption{Instagram mobile application interface.}
    \label{fig:insta-example}
\end{figure}

We gathered profile information of all public users (5,170,062 in total) as well as all of their publications (known as ``{feeds}'' on Instagram) for a three-year period between December 2011 and December 2014. There were 153,979,348 data objects called ``media'', which include a picture or a video along with some metadata such as hashtags, caption, timestamp, and URLs. This paper only focuses on pictures, which takes up a large majority (97\%) of all media on Instagram. Figure~\ref{fig:insta-example} shows an example profile and feed, where profile includes user-level counts (e.g., posts, followers and following) and feed includes images and picture-level metadata (e.g., likes, caption, hashtags, comments and geolocation, if any). All of these pieces of information can be accessed through Instagram's Application Programming Interface (API).

One important aspect considered in this paper is geography of selfie users, which were inferred by mapping geolocation tags in the photo content. Instagram is known to have high rates of photos that contain geotags. Among all media gathered, 35,030,356 pictures published by 770,095 users contained any geolocation information. The Global Administrative Areas database~\cite{gadm} was used to map location coordinates to corresponding country and city names. 

Another aspect considered is user demographics, which we inferred from photos of users by the Face++ tool~\cite{face++}. Face++ is  an online API that detects faces in a given photo and predicts information about each person in the photo such as age and gender. Its accuracy is known to be over 90\%~\cite{facesengageus}. Age is given in years along with a confidence range; gender is given as `male' or `female' with a confidence value between 0 and 100; and smile is given as a score between 0 and 100. We ran Face++ for a random subset of photos and gained demographic information for 2,286,401 pictures posted by 738,901 distinct users. 

\subsection{Data Validation}
To understand potential bias in data, we compare statistics obtained from our data with those of other reports on Instagram. The service reached 300 million active users in 2014 with more than 30 billion photos shared on the network.\footnote{{http://instagram.com/press/}} 
A research conducted by Pew showed that Instagram is not only increasing its overall user base, but also is seeing a significant growth in almost every demographic group in the United States. Most notably, 53\% of young adults between age 18 and 29 used the service in 2014, compared to 37\% a year before. The service is also known to have more female users than males~\cite{pewreport2014}.

\begin{figure}[h]
    \centering
    \hspace*{-5mm}
    \includegraphics[width=0.525\textwidth]{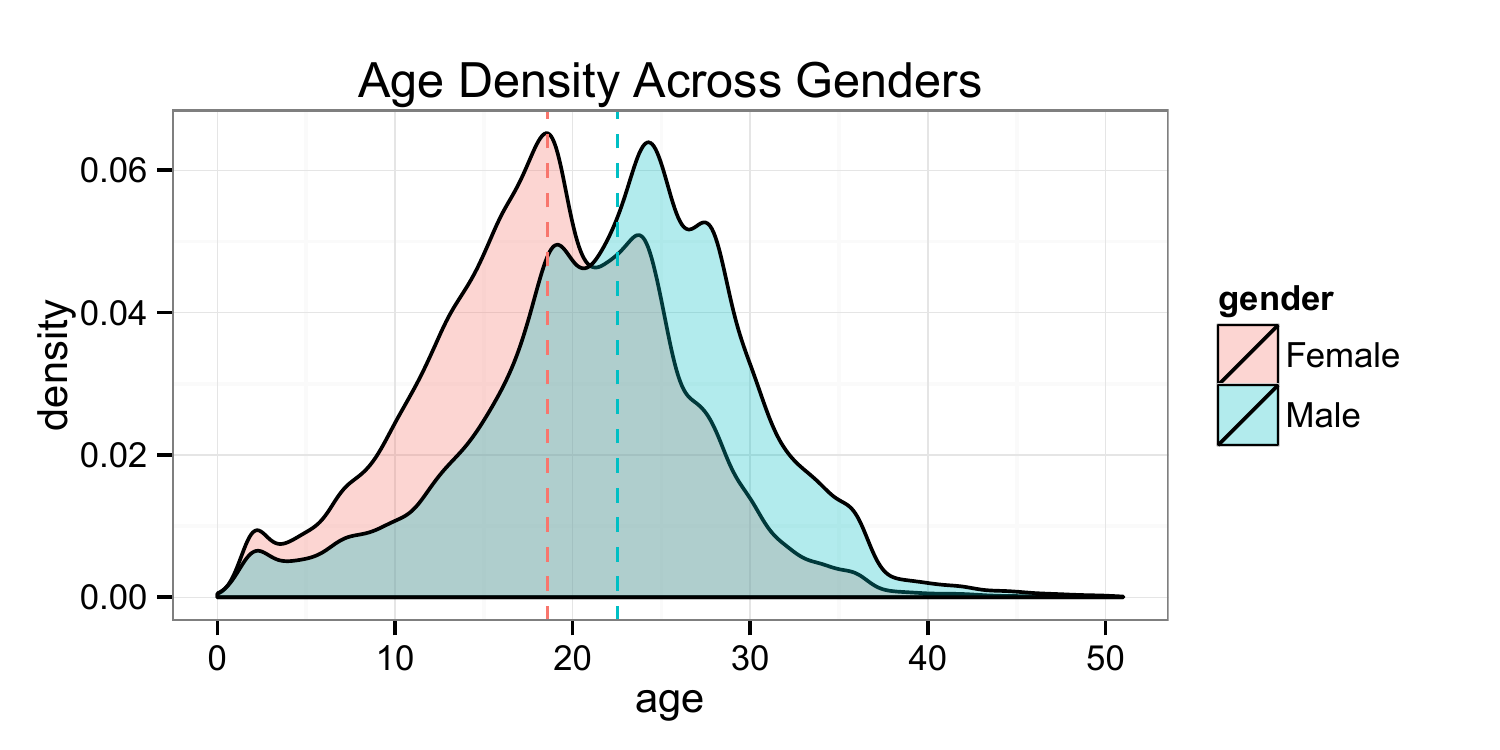}
    \vspace*{-3mm}
    \caption{Density plot of users ages, separated by gender.}
    \label{fig:genderandage}
\end{figure}

Given its massive scale, findings in this study are bound to insights from a small subset   of data. Nonetheless, data we observed had similar properties to what was reported on Instagram. We examined the age and gender distribution of users in our data. We selected a random sample of 100,000 users with at least 10 pictures and examined the profile pictures of such users. The resulting age and gender distribution is shown in Figure~\ref{fig:genderandage}, where 62\% of the sample users are inferred as female and the median ages are 18 and 23 for females and males, respectively. The proportions of different age groups are similar to other reports, like the Pew research and the Selfiecity project~\cite{tifentaleselfiecity}. 

\begin{figure*}[t]
	\centering
	\begin{subfigure}{.22\textwidth}
        \includegraphics[width=\textwidth]{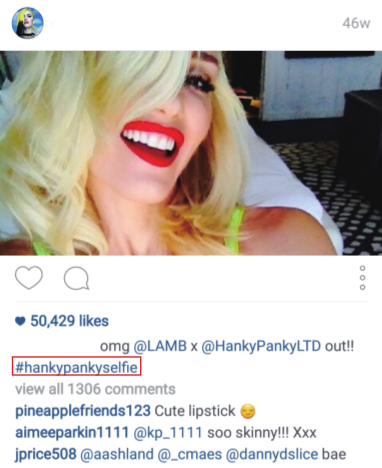}
        \caption{\textsf{Selfie}}\label{subfig:es}
	\end{subfigure}\hspace{0.025\textwidth}
	\begin{subfigure}{.22\textwidth}
        \includegraphics[width=\textwidth]{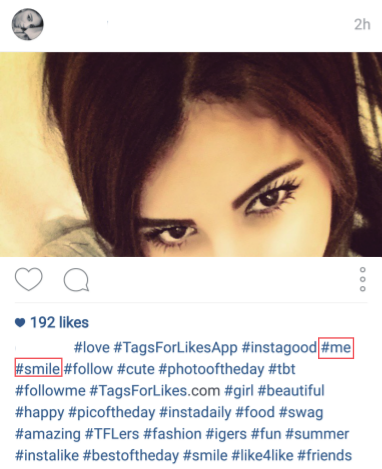}
        \caption{\textsf{Alt}}\label{subfig:ea}
	\end{subfigure}\hspace{0.025\textwidth}
	\begin{subfigure}{.22\textwidth}
        \includegraphics[width=\textwidth]{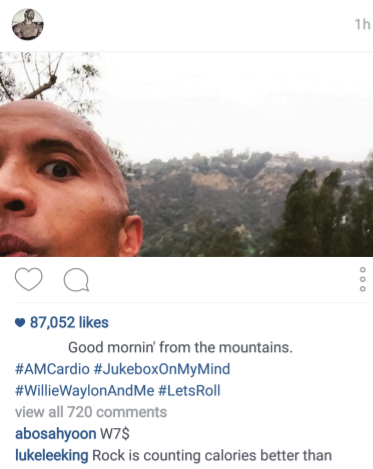}
        \caption{\textsf{Face}}	\label{subfig:ef}
	\end{subfigure}\hspace{0.025\textwidth}
	\begin{subfigure}{.22\textwidth}
        \includegraphics[width=\textwidth]{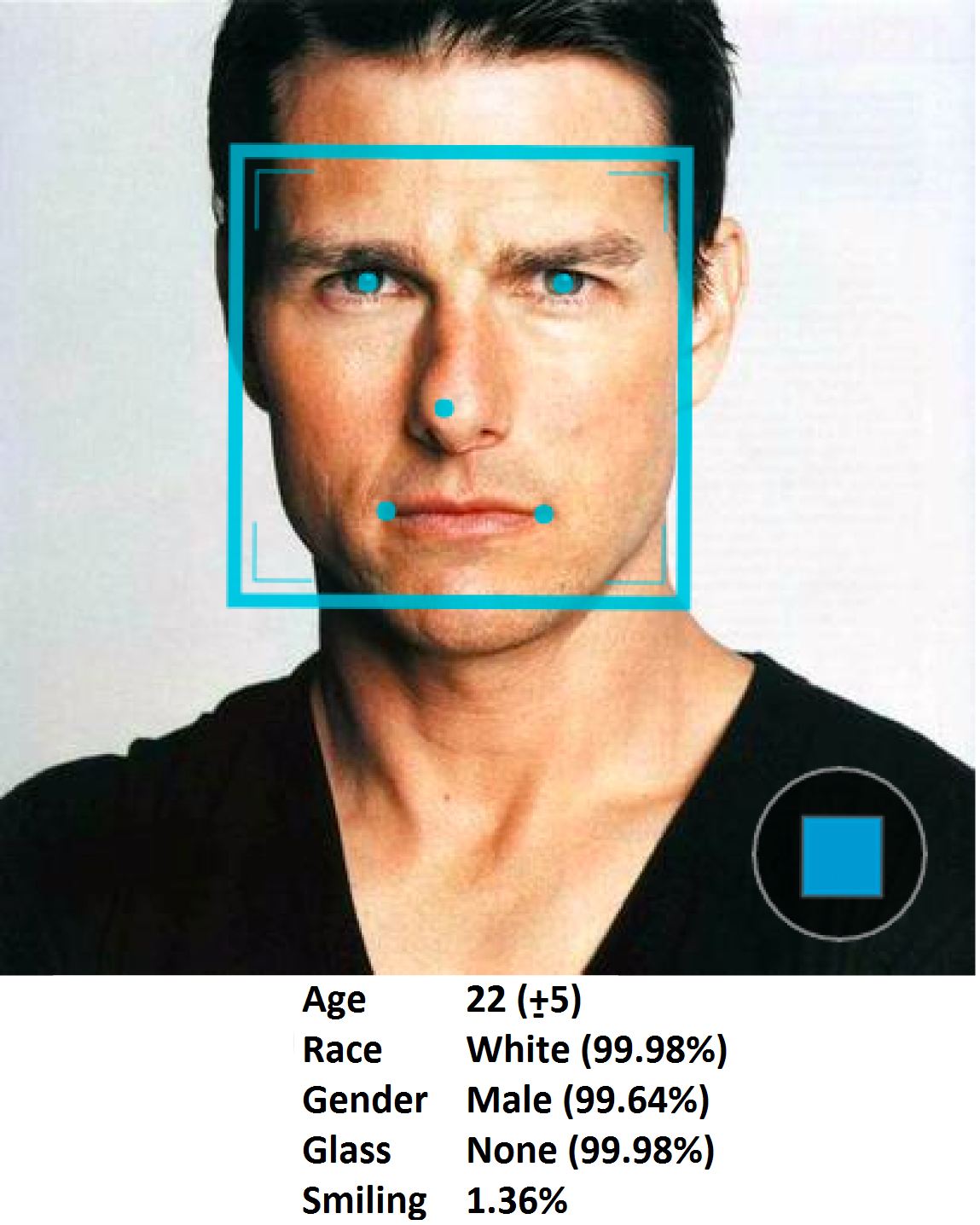}
        \caption{Face++ Inference}	\label{subfig:eff}
	\end{subfigure}
	\caption{Example pictures of \textsf{Selfie}, \textsf{Alt}, and \textsf{Face} datasets as well as features predicted by Face++.}
   \label{fig:exampic}
\end{figure*}

\begin{table*}[t]
{
	\centering
	\begin{tabular}{lp{0.6\textwidth}rrr}
	\toprule
	Dataset &Description &\# Pictures &\# Users\\
	\midrule
	\textsf{Selfie}&Pictures with hashtags containing `selfie' (e.g., \texttt{\#selfie}, \texttt{\#selfietoday})&$1,196,080$ & 214,656\\
	\textsf{Alt}&Pictures with alternative hashtags for `selfie' (e.g., \texttt{\#selca}, \texttt{\#selstagram}) &$2,453,749$ & 242,650\\
	\textsf{Face}&Pictures with face(s) detected using the Face++ tool& 1,921,207 &315,751\\
	\textsf{All}&Randomly chosen set of pictures & $10,000,019$ &  184,615  \\
	\bottomrule
	\end{tabular}
	\caption{Number of media and users in each of the four datasets used in this paper.}
	\label{tab:datasets}
}
\end{table*}

\subsection{Extracting Selfies}
We devised three methods to extract selfie posts. Photos with selfie-related tags indicate what Instagram users identify explicitly as selfies. In addition to two datasets found in this manner, we also examine pictures with faces in general. Note that not all photos belonging to this category are selfies (i.e., photos taken of oneself), yet the third dataset will help us understand the engaging effects of faces. Lastly we utilized a random set of photos for comparison. The summary of the four datasets used in the remainder of the paper follows: 

\begin{enumerate}
    \item \textsf{Selfie}: a collection of pictures that contain the word `selfie' in hashtags. Examples include \texttt{\#selfie}, \texttt{\#selfietime}, and \texttt{\#selfiesunday}, which are an explicit indicator.
    \item \textsf{Alt}: a collection of pictures that include hashtags related to selfie but use variations of the word. For instance, `selca' can be used instead of `selfie' in some contexts. 
    \item \textsf{Face}: a collection of pictures containing one or more faces detected using the Face++ tool. 
    \item \textsf{All}: a random collection of 10M Instagram pictures. We compared it with the other three datasets to identify the distinct characteristic of selfies.
\end{enumerate}

Photos in Figure~\ref{fig:exampic} are examples of the three datasets, which were all posted by popular users on Instagram. Figure~\ref{fig:exampic}(a) is classified as \textsf{Selfie} due to its hashtag \texttt{\#hankypankyselfie}, whereas Figure~\ref{fig:exampic}(b) is classified as \textsf{Alt} for its hashtag 
\texttt{\#me} and \texttt{\#smile}. The face photo in Figure~\ref{fig:exampic}(c) did not contain any selfie-related hashtags, hence it was classified as \textsf{Face} by the Face++ tool. Note that all three types of photos are valid selfie content, which we consider in this paper. Figure~\ref{fig:exampic}(d) shows features detected by Face++ on a celebrity photo of Tom Cruise. Table~\ref{tab:datasets} summarizes the description and quantity (the number of pictures and distinct users) of three selfie datasets as well as that of \textsf{All}. 

Now we describe our heuristic method to identify \textsf{Alt} photos. For this we first need to examine what users call as selfies on Instagram. The \textsf{Selfie} dataset involved a total of 43,874 distinct hashtags containing the word `selfie'. To find  alternative hashtags for `selfie', we calculated a similarity score for each hashtag in a way akin to Pointwise Mutual Information~\cite{PMI}. First, we separated all pictures into two sets: one set containing pictures that either have a single-person face or the hashtag \texttt{\#selfie} (called True or $T$) and another set containing pictures that neither have a face nor the hashtag \texttt{\#selfie} (called Unknown or $U$). The similarity score was then designed in an approximate manner to give higher scores to hashtags in the first set, $T$, as follows:
\begin{equation}
    S_{h} = \dfrac{f_{h,T} {\times} u_{h,T}}{f_{h,U} {\times} u_{h,U}} 
\end{equation}
where $S_{h}$ is the similarity score for a hashtag $h$ in relation to selfie posts. $f_{h,[T,U]}$ is the frequency of the hashtag $h$ in the set $T$ or the set $U$ and $u_{h,[T,U]}$ is the number of users who use the hashtag $h$ in $T$ or $U$. 

\begin{figure*}[t]
	\centering
	\hspace*{-7mm}
        \begin{subfigure}{.26\textwidth}
            \includegraphics[width=\textwidth]{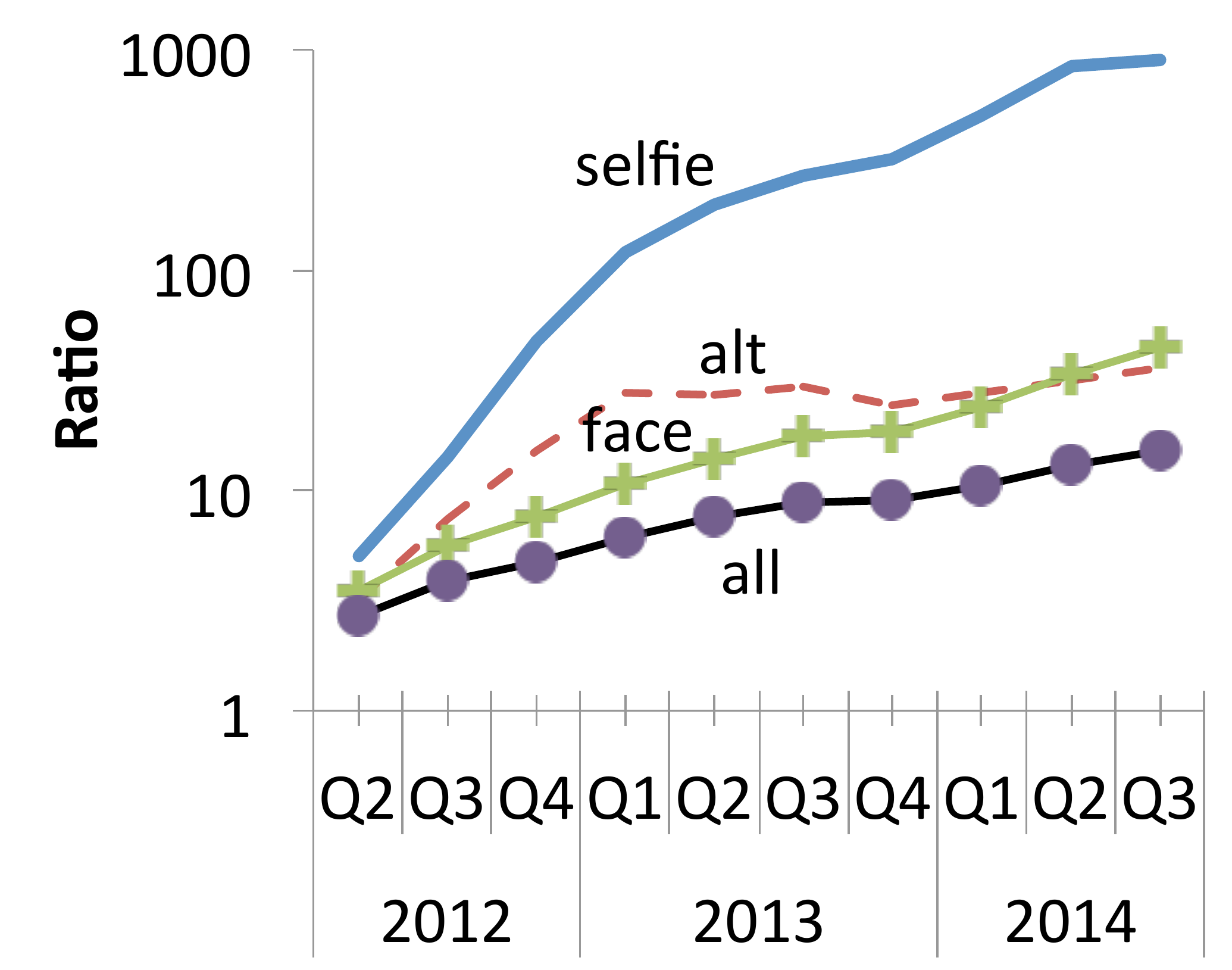}
            \caption{Post frequency}\label{subfig:neu}
        \end{subfigure} 
        \hspace{-0.025\textwidth}
        \begin{subfigure}{.26\textwidth}
            \includegraphics[width=\textwidth]{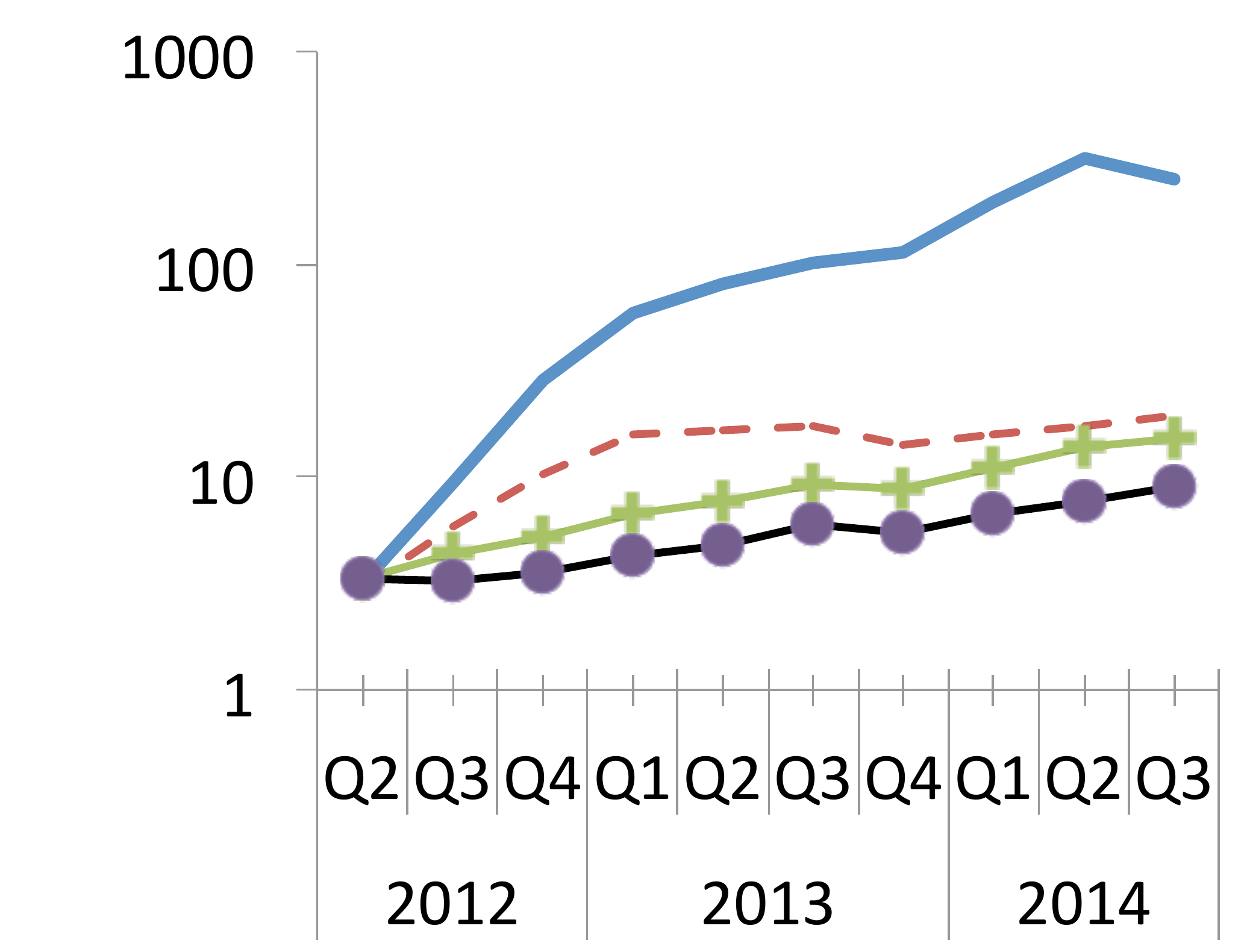}
            \caption{Participating users}\label{subfig:neip}
        \end{subfigure} 
        \hspace{-0.025\textwidth}
        \begin{subfigure}{.26\textwidth}
            \includegraphics[width=\textwidth]{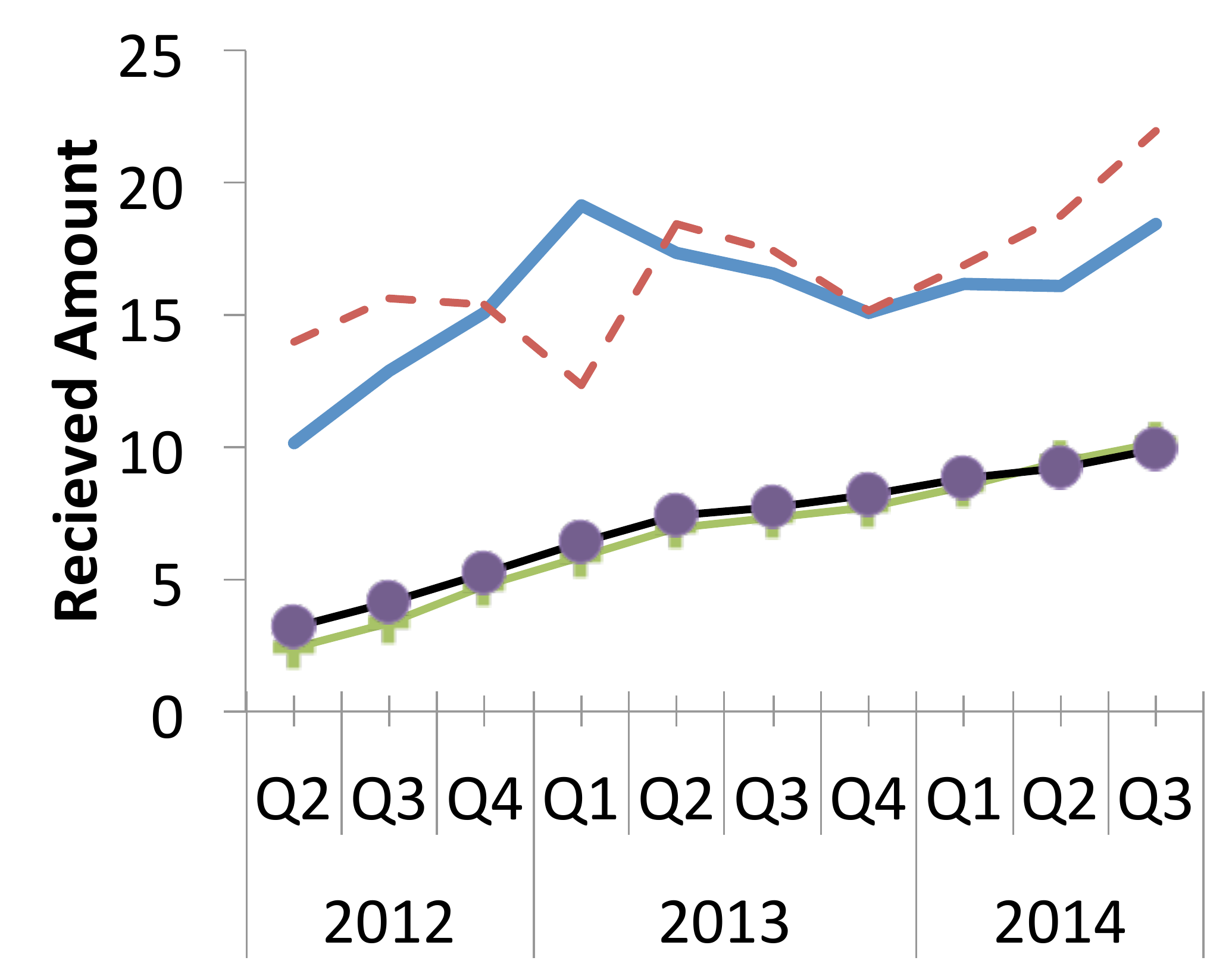}
            \caption{Likes received}	\label{subfig:nel}
        \end{subfigure} 
        \hspace{-0.025\textwidth}
        \begin{subfigure}{.26\textwidth}
            \includegraphics[width=\textwidth]{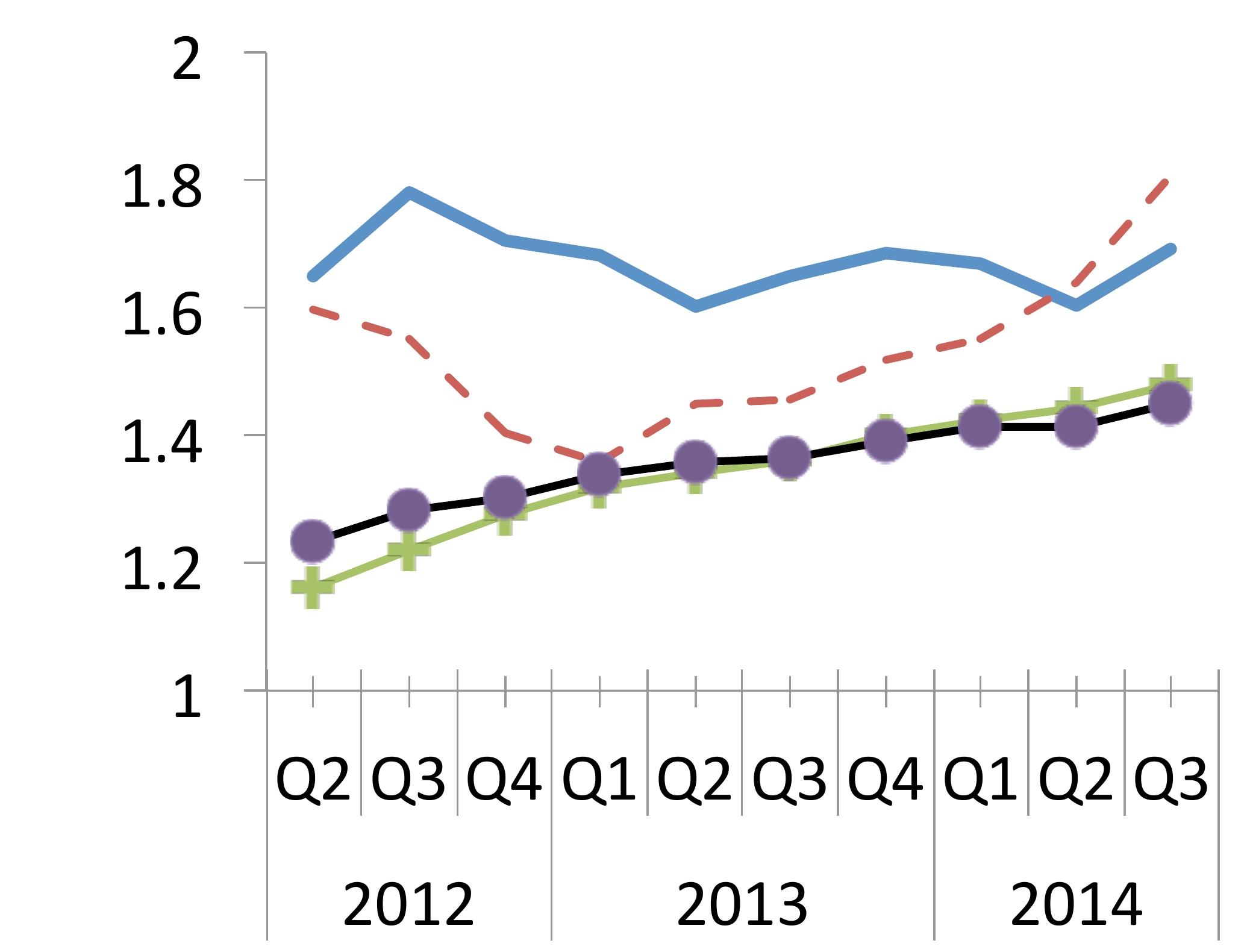}
            \caption{Comments received}	\label{subfig:nec}
        \end{subfigure}
	\hspace*{-7mm}	
        \begin{subfigure}{.26\textwidth}
            \includegraphics[width=\textwidth]{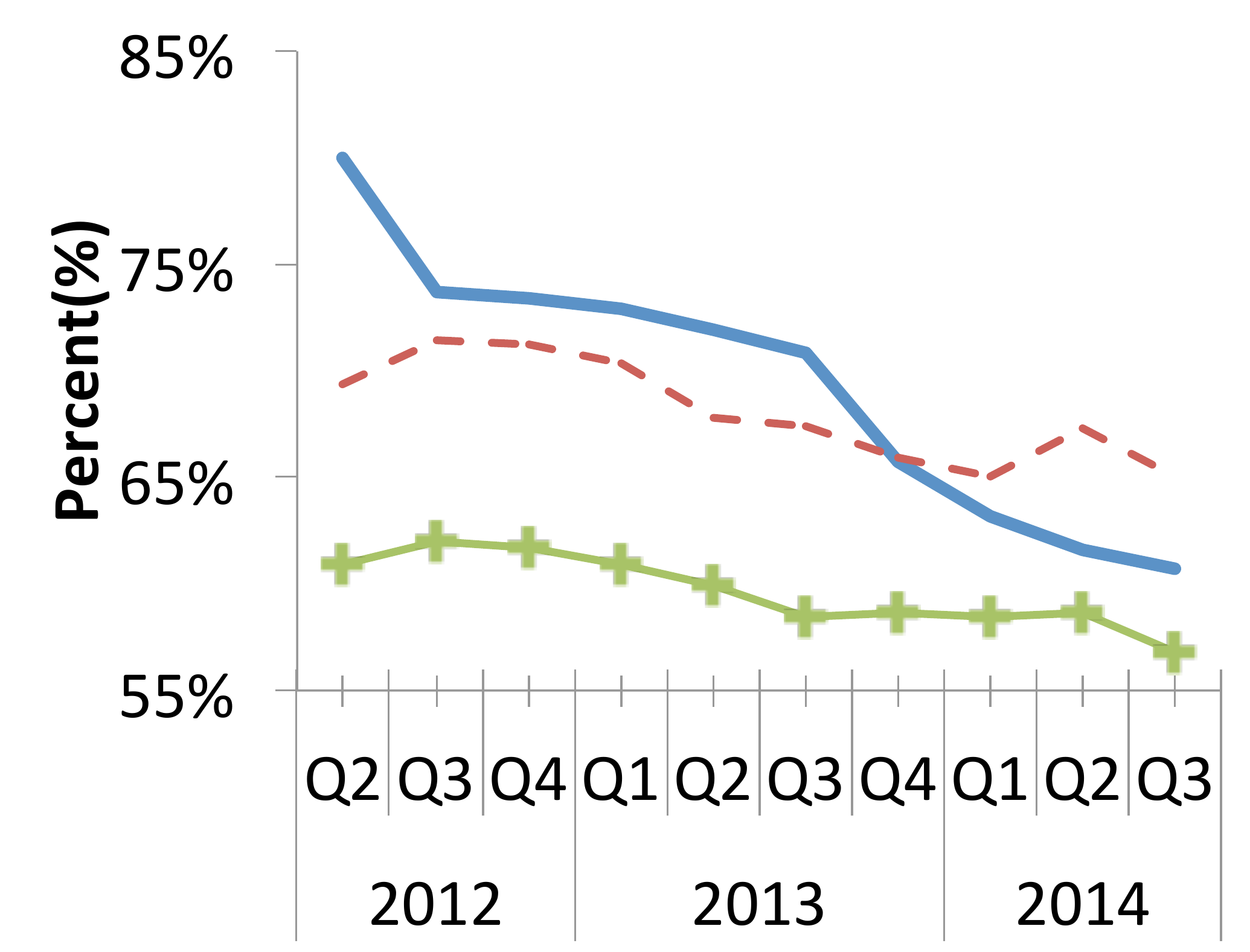}
            \caption{Gender (Female Ratio)}\label{subfig:nef}
        \end{subfigure} 
        \hspace{-0.025\textwidth}
        \begin{subfigure}{.26\textwidth}
            \includegraphics[width=\textwidth]{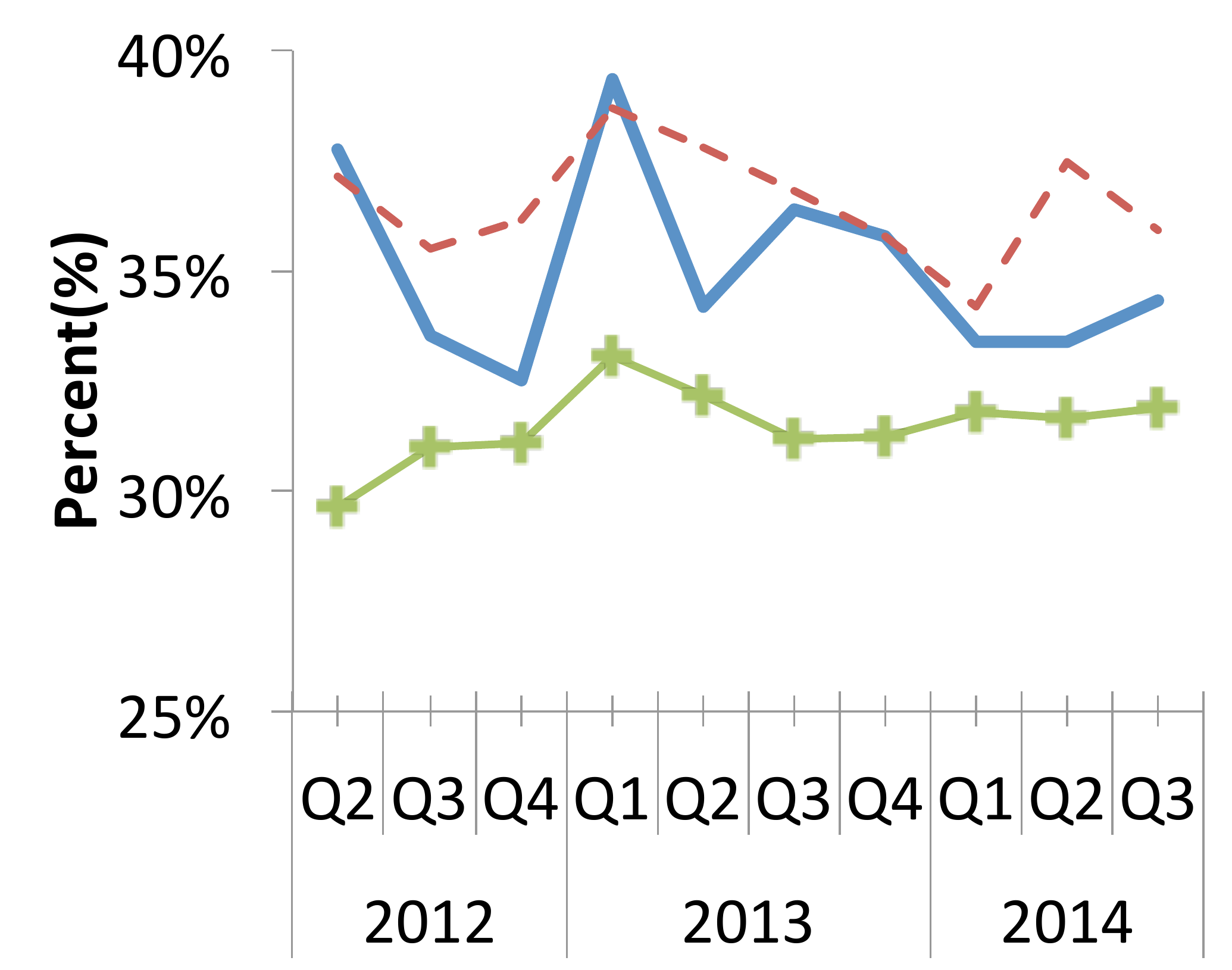}
            \caption{Age (10--19 year olds)}
            \label{subfig:ne10}
        \end{subfigure} 
        \hspace{-0.025\textwidth}
        \begin{subfigure}{.26\textwidth}
            \includegraphics[width=\textwidth]{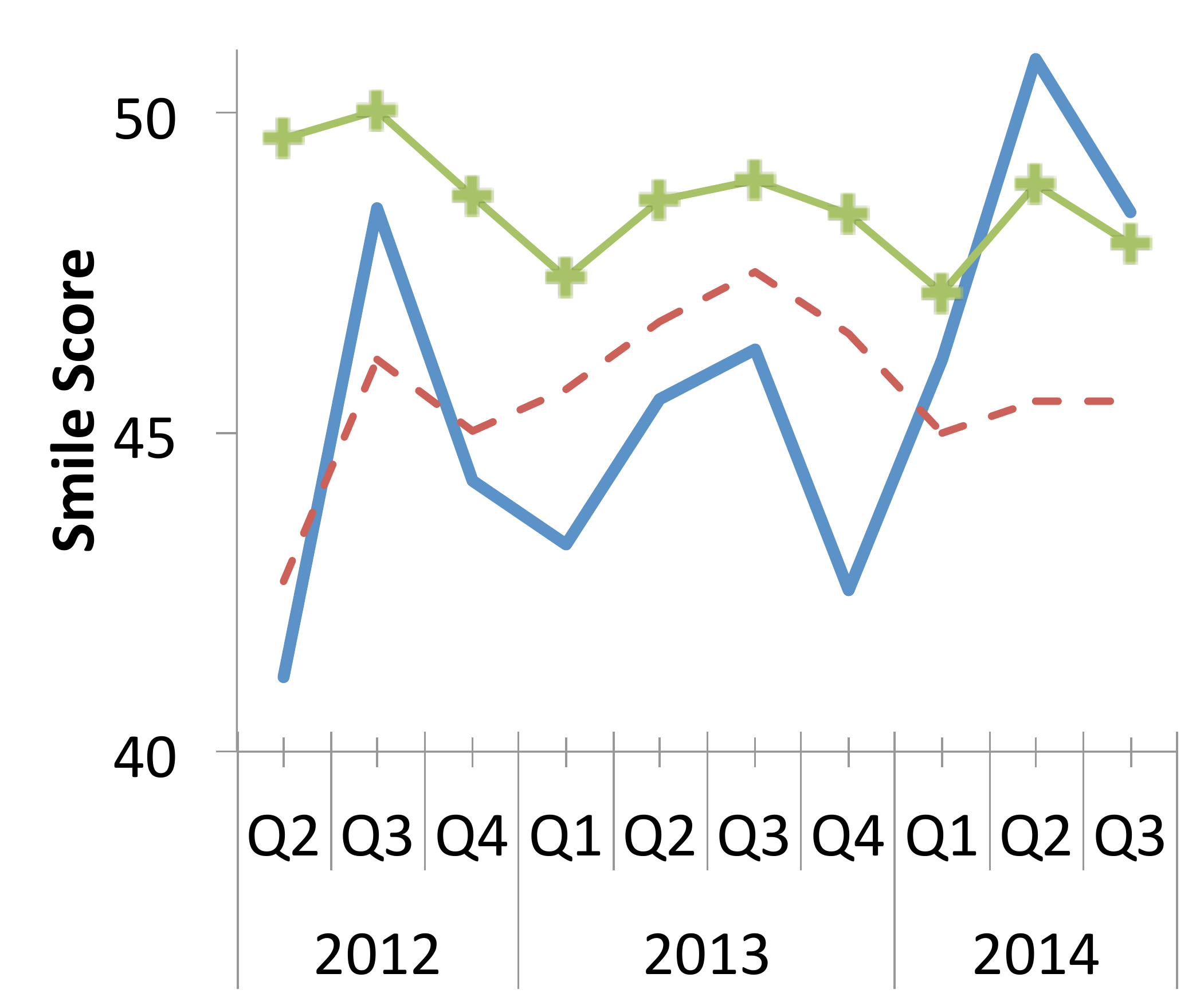}
            \caption{Smile scores}\label{subfig:smile}
        \end{subfigure} 
        \hspace{-0.025\textwidth}
        \begin{subfigure}{.26\textwidth}
            \includegraphics[width=\textwidth]{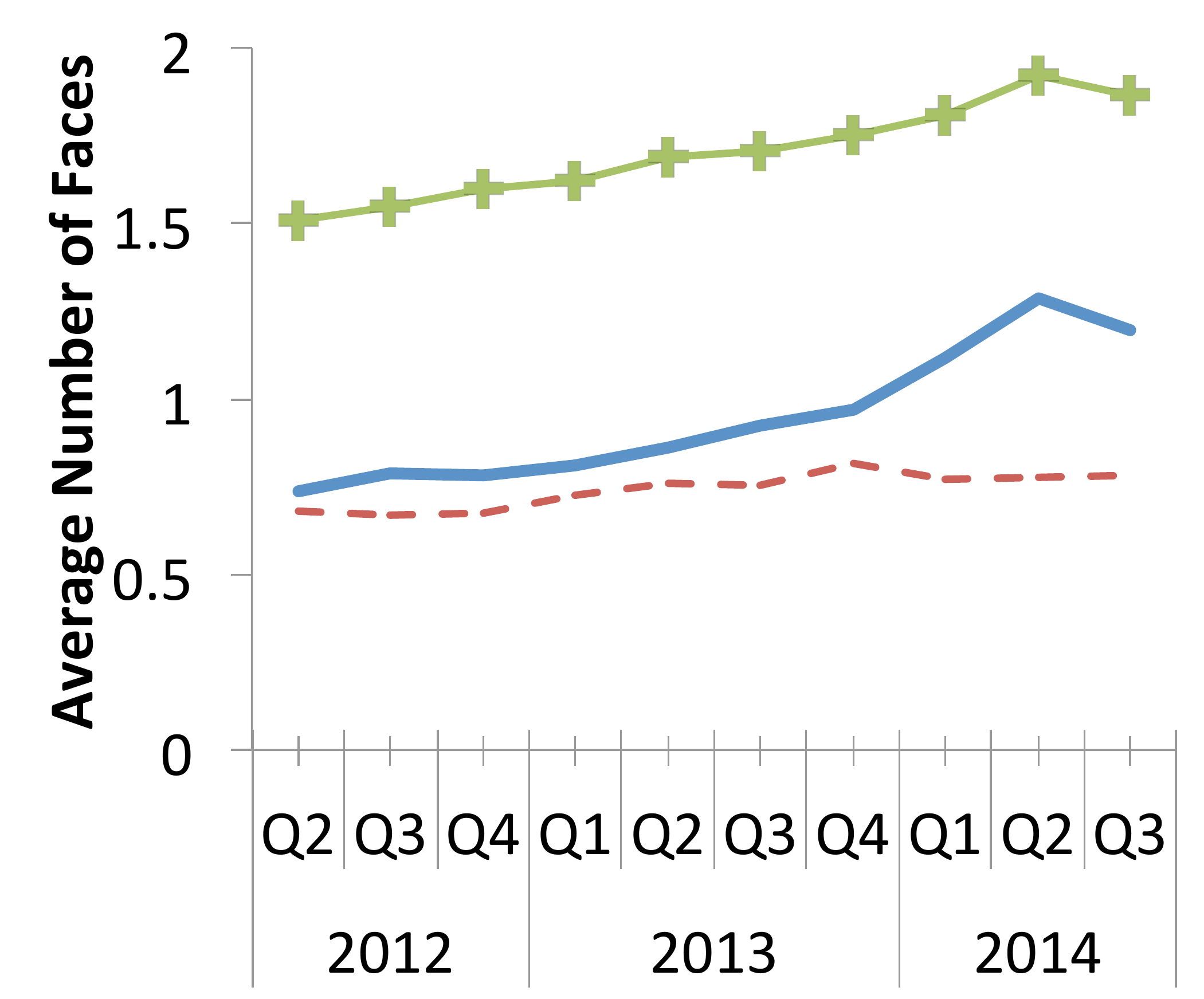}
            \caption{Face counts}
            \label{subfig:anf}
        \end{subfigure}
    \caption{Longitudinal trend of selfie posts across the datasets.}
   \label{fig:evolution}
\end{figure*}

In this manner, we were able to identify words that describe selfies in various languages such as Turkish, Russian, Malaysian, Indonesian, Filipino, etc. Note that the obtained hashtags had high potential to appear with other selfie-related terms, yet they do not cover a complete set of selfie-related terms. The top-10 variant hashtags found with this method were: \texttt{\#shamelessselefie}, \texttt{\#gaybeard}, \texttt{\#butfirst}, \texttt{\#\foreignlanguage{turkish}{özçekim}} (the word representing selfie in Turkish), \texttt{\#ethanymotagiveaway}, \texttt{\#gaysian}, \texttt{\#\foreignlanguage{russian}{лифтолук}} (Russian), \texttt{\#dolledup}, \texttt{\#ozcekim} (Turkish), as well as \texttt{\#pacute} (Malaysian and Filipino). We also included words that are used to describe selfies such as \texttt{\#me} and \texttt{\#self} in the \textsf{Alt} dataset. The final \textsf{Alt} dataset contained a total of 81 variant hashtags.

When we examined the photo content through Face++, the three selfie datasets varied slightly in terms of user demographics. First, the median age of the users in photos were 22, 20, and 21 for \textsf{Selfie}, \textsf{Alt}, and \textsf{Face} respectively. \textsf{Alt} photos contained the youngest users. The proportion of females to males varied from 64\%, 69\%, and 59\% for the three datasets (in same order as above). Photos in \textsf{Alt} were more likely to contain faces of female users, while \textsf{Face} had a better balance of male and female users. Finally, we compared how many faces appear in a given photo, as sometimes multiple faces may appear in a photo (i.e., groupie). The three datasets contained on average 1.12, 0.76, and 1.75 number of faces for the three datasets (in the same order as above). Some photos in \textsf{Alt} were selfies of pets or body parts, in which case they contained zero human face. These variations, although not prominent, may indicate the differences in base demographics.

\section{Temporal Dynamics}

The longitudinal data provides a unique opportunity to examine post and interaction trends  from 2012 to 2014. 

\subsection{Patterns of Selfies Over Time}
We first study how a wide range of features changed over time. We examine post frequency, attention (likes and comments), demographics (gender and age), and image (smile score and number of faces in a given photo) for different definitions of selfies. Demographics and image features depend on Face++ data, thus the \textsf{All} dataset was excluded from these analyses.

\subsubsection{Post Frequency}
Post frequency measures how popular a given photo type is over time. Figure~\ref{fig:evolution}(a) shows the post frequency trends over time, where the $x$-axis represents time in quarters (i.e., three-month periods) and the $y$-axis represents the relative increment or decrement compared to the initial quarter (i.e., the first quarter of 2012). Therefore, a value of 1.0 in this figure means the post volume is identical to what was measured in the initial quarter and a value of 10.0 means an increment by 10 times. 

While the frequency of \textsf{All} increased 15 times (from 103,520 in the first quarter of 2012 to 1,560,697 in the third quarter of 2014), the post frequency of \textsf{Selfie} increased rapidly by 900 times (from 297 to 269,454) over the same time period. \textsf{Alt} and \textsf{Face} also became popular compared to \textsf{All}, yet not at the same degree as \textsf{Selfie}. When we compare the speed, \textsf{All} and \textsf{Face} show a relatively steady growth in volume, whereas the growth of \textsf{Selfie} and \textsf{Alt} is rapid at first and becomes stagnant towards the end of 2014. A similar trend is seen in the graph of participating users who post selfies in Figure~\ref{fig:evolution}(b). \textsf{Selfie} again shows orders of magnitude larger growth than any other content type. \textsf{Selfie} and \textsf{Alt} show a stagnant growth towards the end of 2014 as opposed to \textsf{All} and \textsf{Face}. These growth trends capture well the rapid rise of selfies on Instagram, which seemed to have peaked between 2012 and 2013. 

\subsubsection{Content Popularity}
The amount of attention a photo gained can be inferred by examining the number of likes and comments. Figure~\ref{fig:evolution}(c) shows the absolute geometric mean of likes per picture, which demonstrates that \textsf{Selfie} and \textsf{Alt} receive nearly 2-3 times more likes than the other content types. This means pictures with an explicit marker about `selfie' grab more attention from audience than merely containing a face in a photo. Examining closely, however, the relative gap between \textsf{Selfie} and \textsf{All} decreases over time from nearly 3.2 times during the thriving initial spread to 1.3 times in 2014.

The geometric mean of comments received in Figure~\ref{fig:evolution}(d) shows a similar trend. Again \textsf{Selfie} and \textsf{Alt} receive 1.1--1.5 times more comments than the other content types, although this gap is decreasing over time. This observation indicates that pictures owning a selfie-related hashtag are effective in grabbing attention, yet their engaging effect becomes less pronounced over the years (perhaps as selfies become widespread and become mundane). 

When compared to the recent literature on the effect of containing faces in pictures, the work in~\cite{facesengageus} demonstrated pictures with faces tend to get 38\% more likes and 32\% more comments compared to other content on Instagram. While we cannot make a direct comparison, our results further highlights that attributing particular hashtags (such as \#selfie) could incur even a higher level of attention than merely posting photo with faces. 

\begin{figure*}[t]
    \centering
    \includegraphics[width=\textwidth]{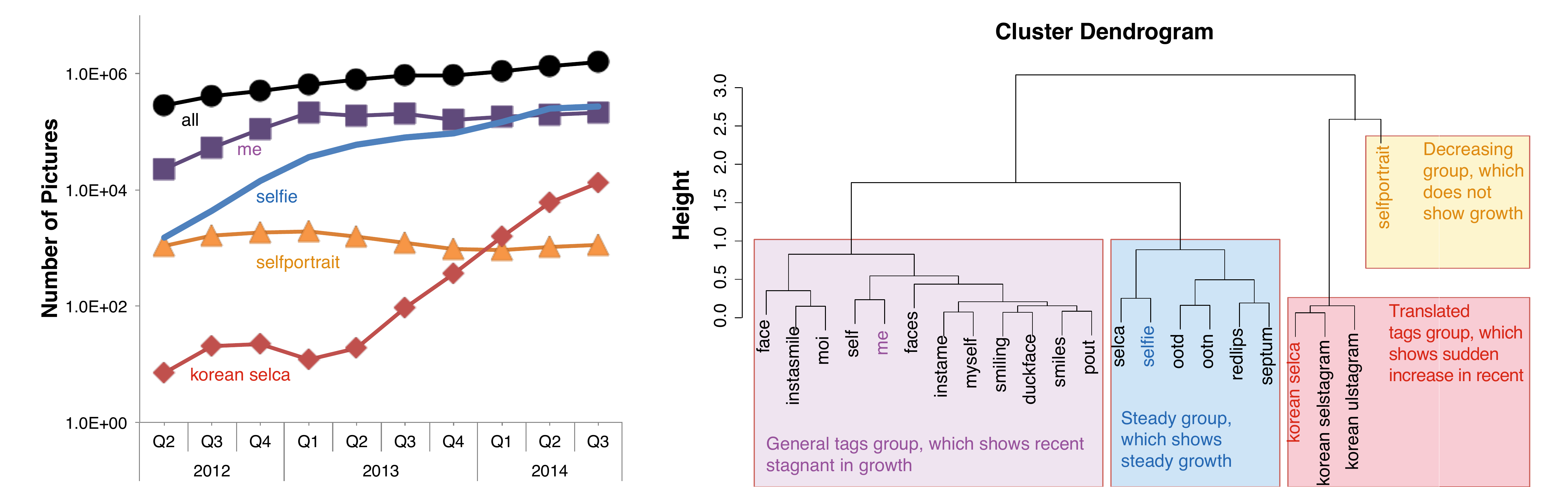}
    \caption{Evolution of tags in \textsf{Alt} and hierarchical clustering based on Pearson correlation.}
    \label{fig:tagevol}
 \end{figure*}

\subsubsection{Demographics}
We next investigate what kinds of users post selfies, by employing the Face++ tool to infer  their age and gender  based on profile pictures. Figure~\ref{fig:evolution}(e) shows the proportion of female users over time. The high female-gender ratio indicates that selfies were initially posted primarily by female users than male users for all three datasets. These rates are high even when we consider the high female prevalence on Instagram. %(60.30\% are females in \textsf{All}). 
During the 3 year period, however, this difference diminished until the ratio almost reached the base gender ratio of the network, as seen by \textsf{Face}.  

Figure~\ref{fig:evolution}(f) shows the demographic makeup for 10--19 year olds. The age distribution  of \textsf{Face} confirms the general perception that young people are the most active users who post photos with faces on Instagram, constituting nearly 32\% of all participating users. Nonetheless, this ratio is even larger for \textsf{Selfie} and \textsf{Alt}, meaning that young people are more likely to tag their pictures with selfie-related hashtags. The gender and age analysis together indicates that young female users on Instagram drove the selfie momentum during the initial stage in 2012, which is indeed confirmed by the plot of percentage of young females over time for each dataset (not included here due to space limitations).

\subsubsection{Smiles and Face Counts}
Would face pictures that are tagged as selfies present more joyful atmosphere? The smile score, detected by Face++, indicates the degree of smile in a face, with a score of 100 indicating the highest level of smile. Comparing the average smile scores in Figure~\ref{fig:evolution}(g), faces in \textsf{Selfie} and \textsf{Alt} are  not more joyful than \textsf{Face}. Overall, the scores of all three datasets are ranged between 40 and 52, which do not necessarily represent a big pleasant smile. We do not see any particular correlation between the smile score and the type of data.

Another question we had was to measure how many faces appear in selfie photos. Would people associate single-person photos as selfies? Figure~\ref{fig:evolution}(h) shows the average number of faces per picture for the three datasets. At a glance, \textsf{Face} contains the highest number of faces (1.5--2.0 faces per picture) than \textsf{Selfie} and \textsf{Alt}. The latter types sometimes included zero faces thereby pushing the average below 1.0, where pictures were on parts of body, pets, or other animals. From mid 2013 and onward, there is a gradual increase in face counts for \textsf{Selfie} dataset, which implies that Instagram users increasingly recognize pictures containing multiple faces as selfies.

\subsection{Trajectory of Selfie Hashtags}
We have so far found similarity between \textsf{Selfie} and \textsf{Alt}, both of which contain hashtags about selfies. It is natural to observe \textit{multiple} variants of the hashtag as they could indicate cultural traits and contexts. In order to observe how different hashtags gained popularity over time, we looked at their adoption trajectory over time. We identified all hashtags in \textsf{Alt} that appeared more than 10,000 times (21 variants) along with \#selfie, then calculated the Pearson's correlation coefficient for growth trajectories of all of these variations. A hierarchical clustering approach was applied to identify hashtags that showed similar growth patterns. 

Figure~\ref{fig:tagevol} shows the result, where hashtags are divided into four groups. The first group, which is the largest in size, contains general description of selfies such as \#me and \#face. Hashtags in this group initially showed high popularity and then became stagnant over time. The second group, which includes social media specific terms such as \#selfie, \#selca and \#ootd (outfit of the day), shows steady growth over time. The third group, containing hashtags in languages other than English, shows a rapid uptake later in time, indicating that the selfie convention has become widely adopted at different times around the world. The last group contains a single hashtag, \mbox{\#selfportrait}, whose growth does not change much over the years. These differences in popularity trajectory of hashtags suggest that underlying mechanisms (i.e., culture, platform) played important roles in how the selfie phenomenon settled.

\begin{figure}[t]
    \centering
    \hspace*{-7mm}
    \begin{subfigure}{.26\textwidth}
        \includegraphics[width=\textwidth]{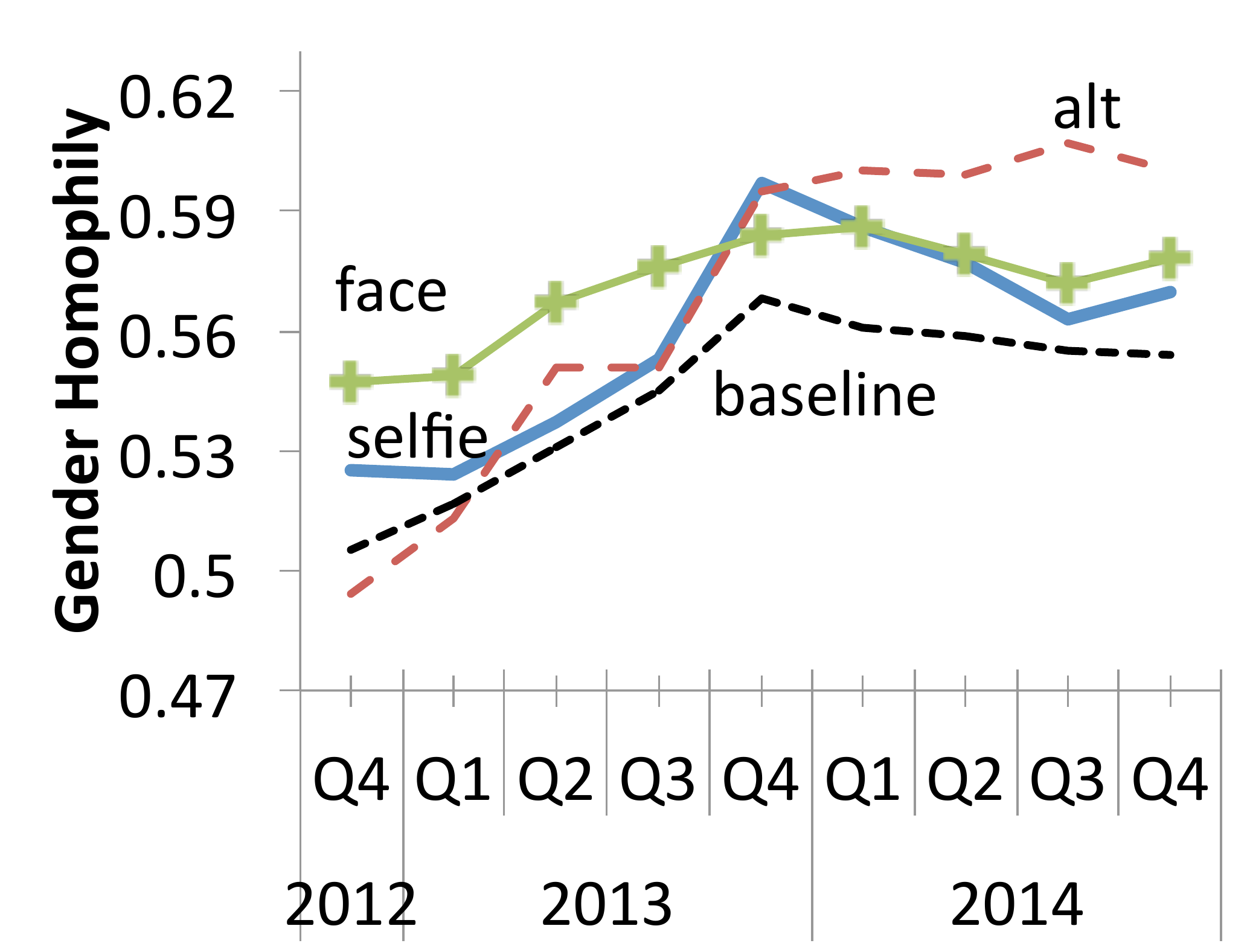}
        \caption{Gender Homophily}\label{subfig:ghs_like}
    \end{subfigure} 
    \hspace{-0.025\textwidth}
    \begin{subfigure}{.26\textwidth}
        \includegraphics[width=\textwidth]{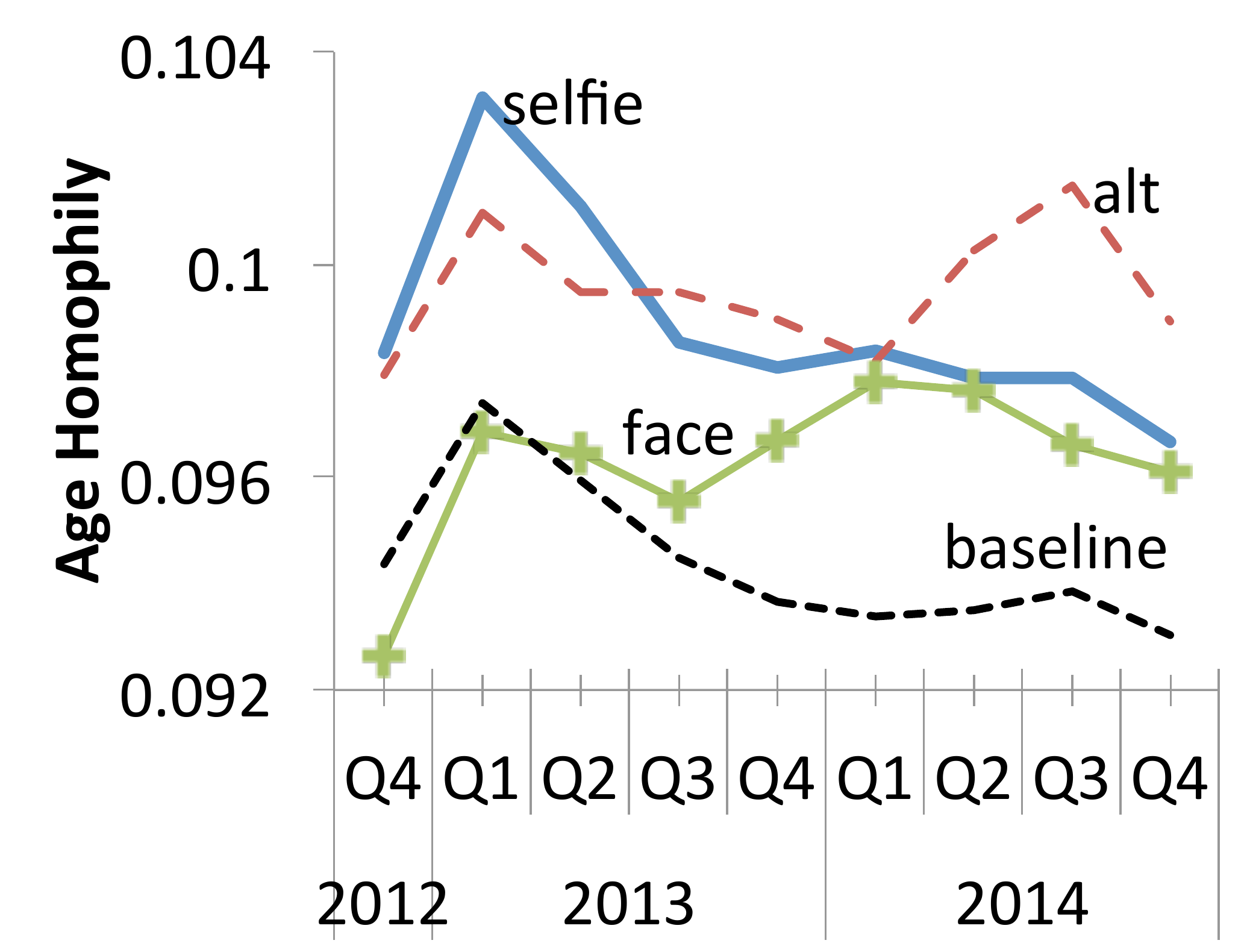}
        \caption{Age Homophily}\label{subfig:ags_like}
    \end{subfigure}
    \caption{Homophily scores for likes. The \textit{baseline} gives the expected scores for random interactions considering the three datasets.}
    \label{fig:homophily_like}
\end{figure}

\begin{figure*}
    \centering
    \includegraphics[width=\textwidth]{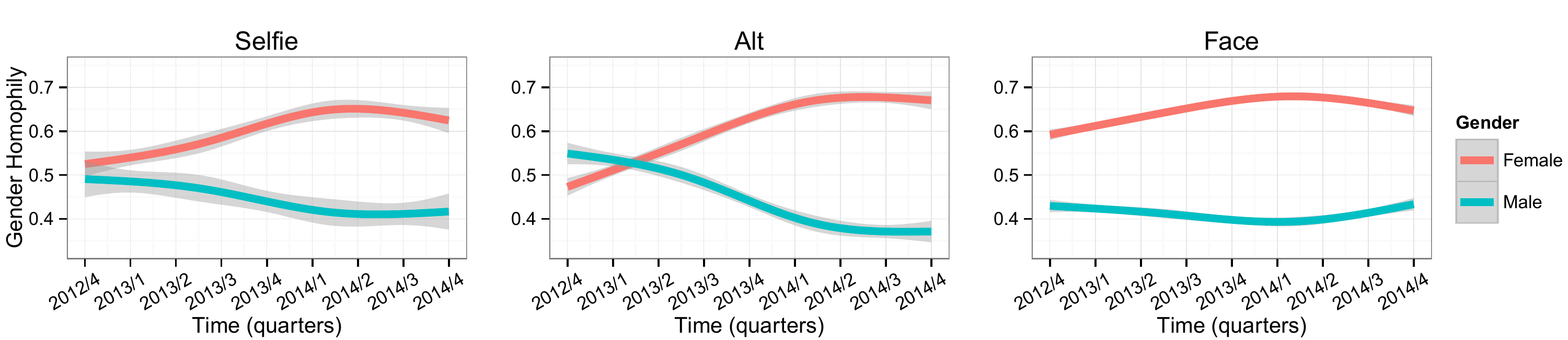}
    \caption{Intra-gender homophily evolution for likes interactions across three datasets:  \textsf{Selfie}, \textsf{Alt}, and \textsf{Face}.}
    \label{fig:intra_gh}
\end{figure*}

\subsection{Selfies as an Interaction Medium}
Since selfies are pictures of people, they represent a structured form of interaction. As shown in the previous subsection, they are an effective medium of communication that incur more likes and comments than other types of content on Instagram. Next, several questions motivate us to examine interaction patterns involving selfies; for instance, how likely males will respond to selfies of other males or other females? Do people tend to interact more frequently with others of the same age? Would these patterns change for pictures explicitly marked as selfies? 

One method to examine these questions is homophily, which describes the tendency of individuals to associate and relate with similar others. Homophily is a central hypothesis that can explain user behaviors in various offline and online social networks~\cite{birdsofafeather}. This study tested homophily by studying the dyadic relationship between selfie owners and their audience based on likes and comments. In particular, we designed the following experiment: first we built three samples, each containing 200,000 randomly chosen pictures from \textsf{Selfie}, \textsf{Alt}, and \textsf{Face} datasets. We then fetched a random set of likes and comments for each of the pictures and determined the age and gender of such interacting users based on Face++. Next, two measures were defined, following the literature on homophily: Gender Homophily (\textit{H\_Gender}) and Age Homophily (\textit{H\_Age}). \textit{H\_Gender} was calculated as follows: 
\begin{equation}
	 \textit{H\_Gender} =  \frac{F_{mm} + F_{ff}}{F_{interactions}}
\end{equation}
where $F_{interactions}$ is the total number of interactions of any kind, $F_{mm}$ is the number of male-male interactions, and $F_{ff}$ is the number of female-female interactions. Hence, \textit{H\_Gender} measures the rate of same-sex interactions out of all combinations seen in data. Its values range from 0 to 1, where 1 represents the highest level of homophily.

Similarly, \textit{H\_Age} was calculated  in the following way:
\begin{equation}
    \textit{H\_Age} = \textrm{RMSE}(A_i, A_j)^{-1}
\end{equation}
where $A_i$ and $A_j$ are lists containing the ages of interacting users and RMSE is the Root Mean Squared Error of two lists, which represents how close they are related. For example, if $\textit{H\_Age} = 0.1$, then $1/\textrm{RMSE}(A_i, A_j) = 0.1$, that is, $\textrm{RMSE}(A_i, A_j) = 10$, which indicates the mean difference in age between interacting users is roughly 10 years. The smaller the differences in age, the higher \textit{H\_Age} will be, indicating greater homophily. We have chosen RMSE instead of mean absolute difference in order to increase the penalty for higher age differences. 

Gender and age homophily scores for likes are shown in Figure~\ref{fig:homophily_like} for each dataset, along with the expected scores for random interactions (labeled as \textit{baseline}) considering the three datasets together. The baseline was built using bootstrap sample in order to get both source and destination users in each quarter, and then calculating the homophily scores for these random pairs of users.

It is possible to observe that \textsf{Alt} and \textsf{Selfie} present more variation in gender homophily scores over the course of three years than \textsf{Face}. To further investigate this finding, we calculated intra-gender homophily scores for likes as follows:
\begin{equation}
    \textit{H\_Male} =  \frac{F_{mm}}{F_{m\_all}}
    \qquad
    \textit{H\_Female} =  \frac{F_{ff}}{F_{f\_all}}
\end{equation}
where $F_{mm}$ and $F_{ff}$ are defined as previously, $F_{m\_all}$ is the total number of interactions of any kind with males, and $F_{f\_all}$ is the total number of interactions of any kind with females. Thus, \textit{H\_Male} and \textit{H\_Female} are proportions of same-sex interactions calculated for each gender separately. The values of each one range from 0 to 1, where 1 represents the highest level of homophily. 

The intra-gender homophily scores of males and females are plotted together for each dataset in Figure~\ref{fig:intra_gh}. Two main facts arise from the graphs. First, there is a clear female bias in likes interactions, which goes in line with female prevalence in the network. Second, both \textsf{Selfie} and \textsf{Alt} present a unique gender homophily evolution. In the beginning, males tend to engage in like interactions more than females, as the scores are close to 0.5 and there are more females in the network. Then over time, their behavior converges to \textsf{Face}'s, suggesting that selfies are becoming more mundane. No significant pattern of gender-level homophily emerged for comments in the datasets. 

In the case of age homophily in likes interactions, the scores for \textsf{Selfie} follow the pattern of \textsf{Alt}'s until the beginning of 2014. After that, \textsf{Selfie} follows the same trend as \textsf{Face}, with decreasing homophily scores, while \textsf{Alt} scores peak and then also decrease. This finding suggests once more that selfies are becoming more widespread. Again, homophily for comments does not present a clear distinction among datasets, although its values are in the same range (0.092 to 0.104) as homophily for likes. The pattern is more pronounced for likes than comments possibly because likes are larger in volume than comments -- likes are a lightweight communication form that happen more often on Instagram than comments.

\section{Cultural Interpretation}

Having examined the longitudinal trends, we now provide explanations for the selfie patterns and examine cultural aspects. We investigate whether the new selfie convention portrays any cultural and socioeconomic contexts (i.e., country-wise variations). This is an important question because other online behaviors have been shown to depend on culture~\cite{culturaldimensions}. To group selfies by cultural boundaries, we aggregated all geotagged pictures by countries and considered only those countries with at least 20 pictures for analysis. The total number of countries analyzed in each dataset is shown in Table~\ref{tab:numcountries} for all indicators used.

\begin{table}[h]
    \centering
    \begin{tabular}{lccc}
      \toprule
      Ind. & \textsf{Selfie} & \textsf{Alt} & \textsf{Face} \\ 
      \midrule
        GGI & 111 & 115  & 117 \\ 
        PV & 54 & 55 & 56 \\ 
        IDV & 67 & 68 & 68 \\ 
        LCS &  53 & 54 & 55  \\ 
        WCS & 53 & 54 & 55 \\ 
        Choice & 54 & 55 & 56  \\ 
        Trust & 54  & 55 & 56 \\ 
        UAI & 67 & 68 & 68 \\ 
      \bottomrule
    \end{tabular}
    \caption{Number of countries per dataset for each indicator.} 
    \label{tab:numcountries}
  \end{table}

As one might expect, selfie patterns differed from one country to another. For instance, the mean age and female-to-male ratio varied as shown in Table~\ref{tab:topbot}, which shows the top-5 and bottom-5 countries based on female prevalence of selfies. South Korea is ranked the top with its 71\% of selfies shared by female users. Even though there is a general bias towards female users that we have demonstrated in the previous section, several countries such as Nigeria and Egypt present a heavy male bias. 

\begin{table}[h]
    \centering
    \hspace*{-1cm}\begin{tabular}{ccc||ccc}
      \toprule
    	& Top 5 & && Bottom 5 & \\
      Country & M.age & F.prev & Country & M.age & F.prev \\ 
      \midrule
      KOR & 16.9 & 0.71 & NGA & 23.5 & 0.31 \\ 
      KAZ & 19.3 & 0.68 & EGY & 22.7 & 0.28 \\ 
      PHL & 17.9 & 0.68 & SAU & 20.4 & 0.28 \\ 
      CHN & 16.6 & 0.67 & KWT & 22.0 & 0.28 \\ 
      UKR & 20.9 & 0.66 & IND & 23.9 & 0.20 \\ 
      \bottomrule
    \end{tabular}\hspace*{-1cm}
    \caption{Top and bottom countries by female prevalence.} 
    \label{tab:topbot}
  \end{table}

In order to test whether these country-wise variations can be explained by cultural contexts, we utilized popular international socioeconomic indicators as well as indicators from two important sources: (i) {World Values Survey (WVS)} that is an individual-level survey probing cultural values of citizens in 59 countries between 2010 and 2014~\cite{worldvaluessurvey} and (ii) {Hofstede's Cultural Dimensions (HCD)} that is a five-dimensional model of cultural differences studied since 1971 by Geert Hofstede~\cite{Hofstede}. 

\subsection{Hypotheses}
We set up three hypotheses that could enrich our understanding of country-wise variations in selfie trends:

\begin{enumerate}
    \item \textbf{Gender Empowerment ($H_1$).}
    There is no consensus on whether selfies enhance male oppression or allows for a way of asserting agency, although the answer is probably more nuanced~\cite{losh}. Nevertheless, it is reasonable to expect that differentiation in gender roles within a country will be reflected in the proportion of women or men taking selfies. We hence hypothesize that women in countries with higher gender equality are more comfortable in sharing selfies publicly than in less equal countries.

    \item \textbf{Self Embellishment \& Membership ($H_2$).}
    If selfies are indeed a manifestation of self embellishment, it is expected that they will be more prevalent in individualistic societies than in collectivist countries~\cite{foster2003}. If, on the contrary, selfies are more widely used as means of belonging and a norm, then they will be more prevalent where citizens feel a strong tie with their local community or with a global connected community. 

    \item \textbf{Intimacy \& Privacy  ($H_3$).}
    If selfies represent one's sense of intimacy and privacy in an online world, then trust in people and the perception of control over one's own life should mediate the behavior. Among relevant socioeconomic indicators, one may consider the level of perceived uncertainty and loss of control. We hypothesize that countries where people are aversive to uncertainty will post comparatively fewer selfies than otherwise.
\end{enumerate}

\subsection{Independent and Control Variables}
To test the first hypothesis, we compared the proportion of females detected by Face++ in each country with several measures of gender equality. Since the proportion of women in each country varies, we calculated the relative increase or decrease of female prevalence against the observed proportion of women in the World Bank data.\footnote{{http://data.worldbank.org/}} We define \mbox{\emph{GenderBias}} as follows: 
\begin{equation}
    \textit{GenderBias} = \textrm{P}_{\textrm{selfies}} - \textrm{P}_{\textrm{census}}
\end{equation}
where $\textrm{P}_{\textrm{census}}$ is the proportion of females observed in a country.

We used two relevant socioeconomic measures: (i) the Gender Gap Index (GGI) and (ii) Patriarchal Values (PV). The former is published yearly by the World Economic Forum and measures the relative gaps between women and men across four key areas: health, education, economy, and politics~\cite{GGR2014}. The score represents how much the gaps has been closed, so a high score means a more equal society. The latter is a scale of four questions from the WVS in which the respondents state whether they agree with values tied to stereo-typical gender roles~\cite{lyness2014}. A high score here means a less equal society in that cultural values are strongly associated with gender inequality.

To test the second hypothesis, we compared the rate of selfie posts at each country, \textit{Prevalence}, as follows: 
\begin{equation}
    \textit{Prevalence} = \log \frac{F_T}{F_{All}}
\end{equation}
where $F_{T}$ is the frequency of posts in dataset $T$ and $F_{All}$ is the set of posts in the \textsf{All} dataset. We used a logarithmic value since the trend is heavy tailed across countries. 

We used three relevant socioeconomic measures: (i) the Individualism score (IDV), (ii) Local Community Score (LCS), and (iii) World Citizen Score (WCS). The first indicator is from Hofstede’s Cultural Dimensions and describes how separated is an individual from larger social groups in a country. The second and third are from a recent work~\cite{vinson2013} and represent the average value of the response to the following propositions: ``I see myself as a part of my local community'' and ``I see myself as a world citizen''. A high score indicates a strong community membership. These scores are proxies for how strongly tied citizens of a country are to their local community as well as to the international community at large. 

To test the third hypothesis, we resorted to the part of WVS that  is used  as an indicator of \emph{generalized trust}, i.e., the trust in people outside one's social circle~\cite{bjornskov2007}. This question is: ``Generally speaking, would you say that most people can be trusted or that you need to be very careful in dealing with people?'' We used the proportion of citizens who agree with the ``Most people can be trusted'' answer as our \emph{Trust} indicator. We selected the following question to probe for the perception of choice and control: ``How much freedom of choice and control do you feel you have over the way your life turns out?'' Responses are situated in a scale from 1 (\emph{no choice at all}) to 10 (\emph{a great deal of choice}), which we averaged per country and used as our \emph{Choice} indicator.

We also selected the dimension \emph{Uncertainty Avoidance} (UAI) from HCD, which indicates a society's tolerance for uncertainty and ambiguity and to what extent the members of a culture feel either uncomfortable or comfortable in unstructured (novel, unknown, unusual) situations. Our hypothesis was that Trust and Choice would be positively correlated and UAI would be negatively correlated to the prevalence of selfies in a country.

Finally, we also considered the following sets of \textit{control} variables to take into account that Instagram is not used evenly across countries.  Although the measures reported in the \emph{Hypotheses} subsection make intuitive sense, they are strongly related to confounding factors that varies either in Instagram or between countries. We chose three variables to control and calculated the \emph{partial correlations} between the variables of interest. Partial correlations allow us to estimate the relationship of two variables $X$ and $Y$ after partialling out the effect of the control variable $Z$. They are equivalent to constructing two linear models using $X$ and $Y$ as dependent variables and $Z$ as independent variable, then correlating the residuals of each linear model. 

The control variables we chose were \emph{log GDP per capita} as an indicator of economic development, \emph{Internet penetration} as a proxy for technology diffusion, and the \emph{average age} of Instagram users (estimated with Face++ data) to account for the different age profiles between countries. Thus, all correlations we report are between the residuals of the variables after their covariance with the control variables has been partialled out. The correlations between measures/indicators and the control variables are not shown here due to space limitations, but are available in our shared repository.

\subsection{Results}
Results are displayed in Table~\ref{tab:cor-cultural}. The positive relationship between GenderBias and gender equality indicators is clear in all datasets, thus confirming $H_1$. This is true both for the equality measured by the country's socioeconomic structure---parity of gender in social living and access to public institutions---as for the cultural values in which the citizens of a country believe. The presence of an effect in all datasets show that this relationship holds for many definitions of selfies. However, if one is not to consider the \textsf{Face} dataset as representing actual selfies, one can argue that this is the consequence of a broader effect of the presence of women in the network. It is interesting to note, however, that the strongest correlations in each of the indicators was not with the \textsf{Face} dataset, but with the \textsf{Alt} dataset.

We could not detect a meaningful relationship between Individualism Score (IDV) and Prevalence for either direction, and Local Community Score (LCS) and World Citizen Score (WCS) show significant correlations in opposite directions. LCS is moderately correlated to selfies tagged as such, which goes in line with the idea that selfies are tied to a sense of belonging to a community, namely the local community. However, this effect seems exclusively related to the \textsf{Selfie} dataset, as the coefficients are negative (although non-significant) in the other datasets. Moreover, WCS is negatively correlated with the \textsf{Alt} dataset and not meaningfully correlated with the other datasets. 

This finding demonstrates a complex relationship between taking selfies and a country's culture of individuation and connectedness. The effect of a country's individualism, if exists, is much smaller than other factors related to belonging to a community, and could not be detected by us. But even these other factors are not easily interpretable and may be related to different conceptions of selfies. The positive relationship  between the \textsf{Selfie} dataset and LCS advocate for the idea that taking selfies---and tagging them as such---is related to the importance a culture gives to belonging to a community. However, we expected that the relationship with WCS would follow the same path, which did not. A possible explanation is that, since the \textsf{Alt} dataset includes many hashtags that represent similar concepts of a selfie in a given country, its negative relationship with WCS spans from the attitude of citizens of the country to adapt and transform foreign ``memes'' into their cultural reality. Thus, a country with citizens that do not strongly identify themselves as world citizens will still have selfies, but adopt different tags. It is worth mentioning that the correlation between the Prevalences of these two datasets (\textsf{Selfie} and \textsf{Alt}) is only moderate ($r = 0.60, p < 0.0001$). 

\begin{table}[t]
    \centering
    \newcolumntype{d}[0]{D{.}{.}{2}}
    \begin{tabular}{llddd}
    \toprule
    Hypotesis: Measure & Ind.   & \multicolumn{1}{c}{\textsf{Selfie}}             & \multicolumn{1}{c}{\textsf{Alt}}                & \multicolumn{1}{c}{\textsf{Face}}               \\
    \midrule
    $H_1:$ GenderBias & GGI    & 0.34^{***}       & 0.41^{***}       & 0.32^{***}     \\
                    & PV     & \texttt{-}0.20^{\diamond} & \texttt{-}0.38^{***}      & \texttt{-}0.19^{\diamond} \\
    \hline
    $H_2:$ Prevalence  & IDV    & 0.09             & 0.06             & \texttt{-}0.04            \\
                       & LCS    & 0.22^{*}             & \texttt{-}0.13            & \texttt{-}0.07            \\
                       & WCS    & 0.08             & \texttt{-}0.17^{\diamond}  & \texttt{-}0.03            \\
    \hline
    $H_3:$ Prevalence  & Choice & \texttt{-}0.04            & \texttt{-}0.19^{\diamond} & 0.13             \\
                       & Trust  & \texttt{-}0.19^{\diamond} & \texttt{-}0.17            & \texttt{-}0.29^{**}       \\
                       & UAI    & 0.14             & 0.21^{*}         & 0.31^{***}       \\
    \bottomrule
    \end{tabular}
\\Stars represent significance values: $p<0.0001 (^{***})$, $p<0.001 (^{**})$,  $p<0.01 (^{*} )$ and $p<0.05 (^{\diamond})$.  
    \caption{Correlations between selfie-related measures and sociocultural measures. There is a complex relationship between taking selfies and a country’s culture. The chance of using selfie-related hashtags was higher for cultures with stronger local community membership as well as weaker perception of privacy. 
}
    \label{tab:cor-cultural}
\end{table}

As for $H_3$, Choice, Trust and Uncertainty Avoidance (UAI) are related to Prevalence, but in the opposite direction that we expected. In countries where the citizens trust each other and feel they have more control over their lives or are not as aversive to uncertain outcomes, people take \emph{fewer} selfies relative to other kinds of pictures. The analysis shows that selfies are not inhibited by a sense of lack of control and certainty but somewhat stimulated by it. We may speculate two (non-excluding) scenarios that could explain our finding:
1) selfies are an assertion of control over one's identity, so they are more important in places where citizens feel they need this;
2) part of what drives selfies is an attitude or set of values that also promotes lack of trust, a sense of lack of control, and aversion to uncertainty. 
Unfortunately, we cannot distinguish these two scenarios from our results.

\section{Discussion and Conclusion}

\subsection{Implications}
Selfies are ever more present in today’s online culture. This work presented a measurement study based on a large amount of data gathered from Instagram, and defined selfies through three different ways to understand the whos, wheres, and hows of its patterns. We investigated the distributions of post frequency, likes, comments, age, gender, smile scores, and face counts over the course of three years. These patterns collectively show that selfies have become extremely popular (i.e., spreading to a wider set of users in terms of number, age, and gender bias). We examined how different variants of the selfie hashtag gained popularity over time. The longitudinal study also explored the role of homophily in terms of age and gender in selfie-oriented lightweight interactions. 

These temporal patterns showed country-wise variations, some of which could be explained by cultural contexts and others need further investigation. This paper showed that gender equality indicators are tightly related to the proportion of women that appear in different definitions of Instagram’s selfies, which goes in line with views that selfies mobilize the power dynamics of representations and promotes empowerment (in this case for women)~\cite{wp:selfiedebates2}. This paper also showed that there is a complex relationship between taking selfies and a country’s culture of individuation and connectedness. Finally, in contrast to our expectation, selfies were less prevalent in more trustful and not risk averse cultures. These findings show general tendency and we do not claim that there is any causal relationship with culture and selfies.

In a recent interview, Instragram founder Kevin Systrom said that ``the selfie is something that didn’t really exist in the same way before Instagram.''\footnote{{http://goo.gl/cpqhhE}} Indeed, selfies take center stage on Instagram, and this work shows that they do so in very specific ways. In quantitatively capturing those ways, we offer two main insights to designers of social-networking platforms. First, the adoption of selfies show a high variability across countries. In countries lacking considerable adoption (because of, e.g., gender issues), designers should think about new ways of encouraging specific segments of the population. Second, it is well known that individuals tend to interact with like-minded others. For selfies on Instagram, however, this tendency is further emphasized. As a result, a filter bubble might well emerge~\cite{filterbubble}:  users become separated from other dissimilar users, effectively isolating them in their own cultural and self-portrait bubbles. In a way similar to what researchers in recommender systems have done~\cite{auralist}, designers should build and integrate new algorithmic solutions that partly counter the ominous consequences of self-portrait bubbles.

\subsection{Limitations}
One limitation of this work is that selfies have different meanings to social media users. Face count varied in that some considered single-person photos as selfies, while others allowed multiple faces to be included. Some explicitly identified photos containing human faces, while others tagged pictures of their pets, animals, personal belongings, as well as body parts as selfies. These examples illustrate the paradigm shift in how people define selfies. The current study tried to capture these diverse meanings by borrowing three different definitions. Nonetheless, our methodology is limited by the use of hashtags and images, as not all selfies will contain such explicit markers.

Another limitation is in the scope of cultural interpretation. Understanding cultural contexts is immensely important, but also very challenging because it is difficult to separate out the complex interplay among socioeconomic factors. This work employed a handful of popular indicators and attempted to provide better explanations for country-wise variations. This, although preliminary, is a meaningful first step towards understanding how a new online phenomenon spread across the world.

\subsection{Future Directions}
Many questions addressed by this work could be investigated in greater detail to highlight possible nuances not captured by the experiments done. For example, an evaluation of how exactly Face++ accuracy is impacted by the particularities of selfies (close-up, distorted or partial views of a face, etc.) could help to know if there are adjustments to be made in this respect; a detailed analysis of usage patterns and spread of Instagram across countries could reveal how local differences affect the overall temporal dynamics found.

Another possible direction would be to dig further into user-level analyses. For instance, a deeper investigation of general differences in users activities in the network could allow to identify how to appropriately take these differences into account when studying interactions among users. A diverse approach could be to verify how selfie-related behaviors vary from user to user or even across different classes of users  (e.g., celebrities and occasional users). This could also bring information about the effect of users characteristics on the engagement associated with the selfies they publish.

Given its multifaceted character, selfies present yet many dimensions not explored in this research. The attachment of selfies with emotions, for example, could be investigated combining different sentiment analysis methods~\cite{sentimentanalysis} based on comments, captions and hashtags related to selfies.

\section*{Acknowledgments}
We thank the reviewers and our shepherd, Emre Kiciman, for providing valuable comments that helped us improve the paper. \\

\noindent This work was partially supported by CAPES, CNPq, FAPEMIG, and the Brazilian National Institute of Science and Technology for the Web -- InWeb. This work was also supported by the IT R\&D program of MSIP/KEIT (R0184-15-1037) and the BK21 Plus Postgraduate Organization for Content Science of Korea. 

\balance
%\bibliographystyle{abbrv}
%\bibliography{7_references}   

\end{document}